\documentclass[graybox]{svmult}

\usepackage{mathptmx}       
\usepackage{helvet}         
\usepackage{courier}        
\usepackage{type1cm}        
\usepackage{makeidx}         
\usepackage{graphicx}        
\usepackage{multicol}        
\usepackage[bottom]{footmisc}

\makeindex             

\begin{document}

\title*{Cosmological Birefringence: an Astrophysical test of Fundamental Physics}
\titlerunning{Cosmological Birefringence}
\author{Sperello di Serego Alighieri}
\authorrunning{S. di Serego Alighieri} 
\institute{Sperello di Serego Alighieri \at INAF - Osservatorio Astrofisico di Arcetri, Largo E. Fermi 5, Firenze, Italy\\
 \email{sperello@arcetri.astro.it}}

\maketitle

\abstract{We review the methods used to test for the existence of cosmological birefringence, i.e. a rotation of the plane of linear polarization for electromagnetic radiation traveling over cosmological distances, which might arise in a number of important contexts involving the violation of fundamental physical principles. The main methods use: (1) the radio polarization of radio galaxies and quasars, (2) the ultraviolet polarization of radio galaxies, and (3) the cosmic microwave background polarization. We discuss the main results obtained so far, the advantages and disadvantages of each method, and future prospects.}

\abstract*{We review the methods used to test the existence of cosmological birefringence, i.e. a rotation of the plane of linear polarization for electromagnetic radiation traveling over cosmological distances, which might arise in a number of important contexts involving the violation of fundamental physical principles. The main methods are: the radio polarization of radio galaxies and quasars, the ultraviolet polarization of radio galaxies, and the cosmic microwave background polarization. We discuss the main results obtained so far, the advantages and disadvantages of each method and future prospects.}

\section{Introduction}
\label{sec:1}
Cosmological birefringence (CB)\footnote{Reference \cite{Nio10} has argued that one should rather use the term optical activity, since birefringence refers to the two different directions of propagation that a given incident ray can take in a medium, depending on the direction of polarization. Although we agree with his comment in principle, we keep here the term that has been commonly adopted for the cosmological polarization rotation.} deals with the possibility of a rotation of the plane of linear polarization for electromagnetic radiation traveling over large distances in the universe. This possibility arises in a variety of important contexts, like the presence of 
a cosmological pseudo-scalar condensate, Lorentz invariance violation and CPT violation, neutrino number asymmetry,  
the Einstein Equivalence Principle (EEP) violation. We refer the reader to \cite{Nio08} \cite{Nio10} for comprehensive reviews of these fundamental physical contexts and discuss here, in chronological order, the astrophysical methods that have been used to test for the existence of CB. 

In general, what is needed to test for CB is a distant source of linearly polarized radiation, for which the orientation of the plane of polarization at the emission is known, and can therefore be compared with the one observed on Earth to see if it has rotated during its journey. Since CB, if it exists, presumably increases with the distance traveled by the radiation, tests should use the most distant sources.

\section{Radio polarization of radio galaxies and quasars}
\label{sec:2}
Reference \cite{Car90} was the first to suggest to use the polarization at radio wavelengths of radio galaxies and quasars to test for CB\footnote{Reference \cite{Bir82} had earlier claimed a substantial anisotropy in the angle between the direction of the radio axis and the direction of linear radio polarization in a sample of high-luminosity classical double radio sources, but used it to infer a rotation of the Universe, not to test for CB.} . It has used the fact that extended radio sources, in particular the more strongly polarized ones, tend to have their plane of integrated radio polarization, corrected for Faraday rotation, usually perpendicular and occasionally parallel to the radio source axis \cite{Cla80}, to put a limit of $6.0^o$ at the 95\% confidence to any rotation of the plane of polarization for the radiation coming from these sources in the redshift interval $0.4<z<1.5$. 

Reanalyzing the same data, \cite{Nod97} claimed to have found a rotation of the plane of polarization, independent of the Faraday one, and correlated with the angular positions and distances to the sources. Such rotation would be as much as 3.0 rad for the most distant sources. However, several authors have independently and convincingly rejected this claim, both for problems with the statistical methods \cite{Eis97, Car97, Lor97}, and by showing that the claimed rotation is not observed for the optical/UV polarization of 2 radio galaxies (see the next section) and for the radio polarization of several newly observed radio galaxies and quasars \cite{War97}. 

In fact, the analysis of \cite{War97} is important also because it introduces a significant improvement to the radio polarization method for the CB test. The problem with this method is to estimate the direction of the polarization at the emission. Since the radio emission in radio galaxies and quasars is due to synchrotron radiation, the alignment of its polarization with the radio axis implies an alignment of the magnetic field. Theory and MHD simulations foresee that the projected magnetic field should be perpendicular to strong gradients in the total radio intensity \cite{Beg84, Sai88}. For example, for a jet of relativistic electrons the magnetic field should be perpendicular to the local jet direction at the edges of the jet and parallel to it where the jet intensity changes \cite{Bri84}. However such alignments are much less clear for the {\it integrated} polarization, because of bends in the jets and because intensity gradients can have any direction in the radio lobes, which emit a large fraction of the polarized radiation in many sources. In fact it is well known that the peaks at $90^o$ and $0^o$ in the distribution of the angle between the direction of the radio polarization and that of the radio axis are very broad and the alignments hold only statistically, but not necessarily for individual sources (see e.g. fig. 1 of \cite{Car90}) . More stringent tests can be carried out using high resolution data and the local magnetic field's alignment for individual sources \cite{War97}, although to our knowledge only once \cite{Lea97} this method has been used to put quantitative limits on the polarization rotation. For example, using the data on the 10 radio galaxies of \cite{Lea97}, reference \cite{Car98} obtains an average constraint on any CB rotation of $\theta = -0.6^o \pm 1.5^o$ at the mean redshift $\left\langle z\right\rangle =0.78$. However the comment by \cite{Lea97} remained unpublished and does not explain convincingly how the angle between the direction of the local intensity gradient and that of the polarization is derived. For example for 3C9, the source with the best accuracy, \cite{Lea97} refers to \cite{Kro96}, who however do not give any measure of local gradients. 

\section{Ultraviolet polarization of radio galaxies}
\label{sec:3}
Another test for CB uses the perpendicularity between the direction of the elongated structure in the ultraviolet (UV)\footnote{When a distant radio galaxy ($z>0.7$) is observed at optical wavelengths ($\lambda_{obs.} \sim 5000 \AA $), these correspond to the UV in the rest frame ($\lambda_{em.} \leq 3000 \AA $).} and the direction of linear UV polarization in distant powerful radio galaxies. The test was first performed by \cite{Cim94, diS95}, who obtained that any rotation of the plane of linear polarization for a radio galaxy at z=2.63 is smaller than $10^o$. 

Although this UV test has sometimes been confused with the one described in the previous section, probably because they both use radio galaxies polarization, it is a completely different and independent test, which hinges on the well established unification scheme for poweful radio-loud AGN \cite{Ant93}. This scheme foresees that powerful radio sources do not emit isotropically, but their strong optical/UV radiation is emitted in two opposite cones, because the bright nucleus is surrounded by an obscuring torus: if our line of sight is within the cones, we see a quasar, otherwise we see a radio galaxy. Therefore powerful radio galaxies have a quasar in their nuclei, whose light can only be seen as light scattered by the interstellar medium of the galaxy. Often, particularly in the UV, this scattered light dominates the extended radiation from radio galaxies, which then appear elongated in the direction of the cones and strongly polarized in the perpendicular direction \cite{diS94}. The axis of the UV elongation must be perpendicular to the direction of linear polarization, because of the scattering mechanism which produces the polarization. Therefore this method can be applied to any single case of distant radio galaxy, which is strongly polarized in the UV, allowing CB tests in many different directions. Another advantage of this method is that it does not require any correction for Faraday rotation, which is considerable at radio wavelengths, but negligible in the UV. 

The method can be applied also to the polarization measured locally at any position in the elongated structures around radio galaxies, which has to be perpendicular to the vector joining the observed position with the nucleus. Using the polarization map in the V-band ($\sim 3000\AA$ rest-frame) of 3C 265, a radio galaxy at z=0.811 \cite{Tra98}, reference \cite{War97} obtain that the mean deviation of the 53 polarization vectors plotted in the map from the perpendicular to a line joining each to the nucleus is $-1.4^o \pm 1.1^o$.
However usually distant radio galaxies are so faint that only the integrated polarization can be measured: strict perpendicularity is expected also in this case, if the emission is dominated by the scattered radiation, as is the case in the UV for the strongly polarized radio galaxies \cite{Ver01}. 

Recently the available data on all radio galaxies with redshift larger than 2 and with the measured degree of linear polarization larger than $5\%$ in the UV (at $\sim 1300 \AA$) have been re-examined, and no rotation within a few degrees in the polarization for any of these 8 radio galaxies has been found \cite{diS10}. Also, assuming that the CB rotation should be the same in every direction, an average constraint on this rotation $\theta = -0.8^o \pm 2.2^o$ ($1\sigma$) at the mean redshift $\left\langle z \right\rangle =2.80$ has been obtained \cite{diS10}.
The same data \cite{diS10} have been used by \cite{Kam10} to set a CB constraint in case of a non-uniform polarization rotation, i.e. a rotation which is not the same in every direction: in this case the variance of any rotation must be $\left\langle \theta ^2 \right\rangle \leq (3.7^o)^2$. It has also been noticed \cite{Mew10} that the CB test using the UV polarization has advantages over the other tests at radio or CMB wavelengths, if CB effects grow with photon energy (the contrary of Faraday rotation), as in a formalism where Lorentz invariance is violated but CPT is conserved \cite{Kos01, Kos02}.

\section{Cosmic microwave background polarization}
\label{sec:4}
The most recent method to test for the existence of CB is the one that uses the Cosmic Microwave Background (CMB) polarization, which is induced by the last Thomson scattering of a decoupling photon at $z\sim 1100$, resulting in a correlation between temperature gradients and polarization \cite{Lep98}. After the first detection of CMB polarization anisotropies by DASI \cite{Kov02}, there have been several CB tests using the CMB polarization pattern. Only the most recent ones are summarized here. The BOOMERanG collaboration, revisiting the limit set from their 2003 flight, find a CB rotation $\theta = -4.3^o \pm 4.1^o$ (68\% CL), assuming uniformity over the whole sky \cite{Pag09}. The QUaD collaboration finds $\theta = 0.64^o \pm 0.50^o$ (stat.) $\pm 0.50^o$ (syst.) (68\% CL) \cite{Bro09}, while using two years of BICEP data one gets $\theta = -2.6^o \pm 1.0^o$ (stat.) $\pm 0.7^o$ (syst.) (68\% CL) \cite{Chi10, Xia10, Tak10}. Combining 7 years of WMAP data and assuming uniformity, a limit to CB rotation $\theta = -1.1^o \pm 1.3^o$ (stat.) $\pm 1.5^o$ (syst.) (68\% CL) has been set, or $-5.0^o < \theta < 2.8^o$ (95\% CL), adding in quadrature statistical and systematic errors \cite{Kom10}. Therefore, although some have claimed to have detected a rotation \cite{Xia10}, the CMB polarization data appear well consistent with a null CB. In principle the CMB polarization pattern can be used to test CB in specific directions. However, because of the extremely small anisotropies in the CMB temperature and polarization, these tests have so far used all-sky averages, assuming uniformity.

\section{Other methods}
\label{sec:5}
Observations of nearby polarized galactic objects could contribute to the CB test, in particular for those cases where polarization measurements can be made with high accuracy and at very high frequencies (useful if CB effects grow with photon energy). Pulsars and supernova remnants emit polarized radiation in a very broad frequency range. For example, hard X-ray polarization observations of the Crab Nebula \cite{Dea08} have been used to set a limit to CB rotation $\theta = -1^o \pm 11^o$ \cite{Mac08}. However this limit is not particularly stringent, both because of the low accuracy of the X-ray polarization measurement, and because of the limited distance to the source. Future more precise X-ray polarization experiments, such as POLARIX \cite{Cos10}, could much improve the situation.

For an issue related to CB, reference \cite{Hut05} provides evidence that the directions of linear polarization  at optical wavelengths for a sample of 355 quasars ($0\leq z\leq 2.4$) are non-uniformly distributed, being systematically different near the North and South Galactic Poles, particularly in some redshift ranges. Such behaviour could not be caused by CB, since a rotation of randomly distributed directions of polarization could produce the observed alignments only with a very contrieved distribution of rotations as a function of distance and position in the sky. Moreover the claim by \cite{Hut05} has not been confirmed by the radio polarization directions on a much larger sample of 4290 flat-spectrum radio sources \cite{Jos07}.

The rotation of the plane of linear polarization can be seen as different propagation speeds for right and left circularly-polarized photons ($\Delta c/c$). The sharpness of the pulses of pulsars in all Stokes parameters can be used to set limits corresponding to $\Delta c/c \leq 10^{-17}$. Similarly the very short duration of gamma ray bursts (GRB) gives limits of the order of $\Delta c/c \leq 10^{-21}$. However the lack of linear polarization rotation discussed in the previous sections can be used to set much tighter limits ($\Delta c/c \leq 10^{-32}$, see ref. \cite{Gol96}).

In a complementary way to the astrophysical tests described in the previous sections, also laboratory experiments can be used to search for a rotation of the plane of linear polarization. These are outside the scope of this review. For example the PVLAS collaboration has found a polarization rotation in the presence of a transverse magnetic field \cite{Zav06}. However this claim has later been refuted by the same group, attributing the rotation to an instrumental artifact \cite{Zav07}.

\section{Comparison of the CB test methods}
\label{sec:6}

Table 1 summarizes the most important limits set on CB rotation with the various methods examined in the previous sections. Only the best and most recent results obtained with each method are listed. For uniformity, all the results for the CB rotation are listed at the 68\% CL ($1\sigma$), except for the first one, which is at the 95\% CL, as given in the original reference \cite{Car90}. In general all the results are consistent with eachother and with a null CB rotation. Even the CMB measurement by BICEP, which apparently shows a non-zero rotation at the $2\sigma$ level, cannot be taken as a firm CB detection, since it has not been confirmed by other more accurate measurements. Nevertheless we cannot but notice that most CB rotations have a negative value, although the most accurate measurement by QUAD has a positive one. 

\begin{table}
\label{tab:1}
\caption{Comparison of CB test methods.}
\begin{tabular}{lccccl}

\noalign{\smallskip}\svhline\noalign{\smallskip}
Method & CB rotation & Distance & Direction & Ref.\\
\hline\noalign{\smallskip}
RG radio pol. & $\vert \theta \vert < 6^o$ & $0.4<z<1.5$ & all-sky (uniformity ass.) & \cite{Car90} \\
RG radio pol. & $\theta = -0.6^o \pm 1.5^o$ & $\left\langle z \right\rangle = 0.78$ & all-sky (uniformity ass.) & \cite{Car98, Lea97} \\
RG UV pol. & $\theta = -1.4^o \pm 1.1^o$ & z = 0.811 & $RA: 176.37^o, Dec: 31.56^o$ & \cite{War97} \\
RG UV pol. & $\theta = -0.8^o \pm 2.2^o$ & $\left\langle z \right\rangle = 2.80$ & all-sky (uniformity ass.) & \cite{diS10} \\
RG UV pol. & $\left\langle \theta ^2 \right\rangle \leq (3.7^o)^2$ & $\left\langle z \right\rangle = 2.80$ & all-sky (stoch. var.) & \cite{Kam10, diS10} \\
CMB pol. BOOMERanG & $\theta = -4.3^o \pm 4.1^o$ & $z \sim 1100$ & all-sky (uniformity ass.) & \cite{Pag09} \\
CMB pol. QUAD & $\theta = 0.64^o \pm 0.71^o$ & $z \sim 1100$ & all-sky (uniformity ass.) & \cite{Bro09} \\
CMB pol. BICEP & $\theta = -2.6^o \pm 1.2^o$ & $z \sim 1100$ & all-sky (uniformity ass.) & \cite{Chi10, Xia10, Tak10} \\
CMB pol. WMAP7 & $\theta = -1.1^o \pm 2.0^o$ & $z \sim 1100$ & all-sky (uniformity ass.) & \cite{Kom10} \\

\noalign{\smallskip}\svhline\noalign{\smallskip}
\end{tabular}

\end{table}

In practice all CB test methods have reached so far an accuracy of the order of $1^o$ and $3\sigma$ upper limits to any rotation of a few degrees. It has been however useful to use different methods since they are complementary in many ways. They cover different wavelength ranges, and, although most CB effects are wavelength independent, the methods at shorter wavelength have an advantage, if CB effects grow with photon energy. They also reach different distances, and the CMB method is unbeatable in this respect. However the relative difference in light travel time between $z=3$ and $z=1100$ is only 16\%.
The radio polarization method, when it uses the integrated polarization, has the disadvantage of not relying on a firm prediction of the polarization orientation at the source, which the other methods have. Also the radio method requires correction for Faraday rotation. All methods can potentially test for a rotation which is not uniform in all directions, although this possibility has not yet been exploited by the CMB methoid, which also cannot see how an eventual rotation depends on distance. 

In the future improvements can be expected for all methods, e.g. by better targeted high resolution radio polarization measurements, by more accurate UV polarization measurements with the coming generation of giant optical telescopes \cite{Gil08, Nel08, Joh08}, and by future CMB polarimeters such as PLANCK and POLARBEAR \cite{Mil09}.
\begin{acknowledgement}
I would like to thank Jan Browne for a useful discussion.
\end{acknowledgement}

\end{document}